\documentclass[twocolumn,showpacs,preprintnumbers,amsmath,amssymb]{revtex4}
\usepackage{bm}
\input{psfig.sty}
\begin{document}

\title{Network Landscape from a Brownian Particle's Perspective}
\author{Haijun Zhou}
\affiliation{Max-Planck-Institute of Colloids and Interfaces,
14424 Potsdam, Germany}
\date{February 11, 2003}
\begin{abstract}
Given a complex biological or social network, how many clusters should it be 
decomposed into?  We define the distance $d_{i,j}$ from node $i$ to node $j$
as the average number of steps a Brownian particle takes to reach $j$ from 
$i$.
Node $j$ is a global attractor of $i$ if $d_{i,j}\leq d_{i,k}$ for any $k$ of
the graph; it is a local attractor of $i$, if $j\in E_i$ (the set of
nearest-neighbors of $i$) and $d_{i,j}\leq d_{i,l}$ for any $l\in E_i$. Based
on the intuition that each node should have a high probability to
be in the same community as its global (local) attractor on the global (local)
scale, we present a simple method to uncover a network's community structure.
This method is applied to several real networks and some discussion on its
possible extensions is made.
\end{abstract}

\maketitle

A complex networked system, such as an organism's metabolic network and 
genetic interaction network, is composed of a large number of interacting 
agents. The complexity of such systems originates partly from the 
heterogeneity in their interaction patterns, aspects of which include the 
small-world \cite{watts1998} and the scale-free properties 
\cite{barabasi1999,faloutsos1999} observed in many social, biological, 
and technological networks \cite{strogatz2001,albert2002,dorogovtsev2002}. 
Given this high degree of complexity, it is necessary to divide a network 
into different subgroups to facilitate the understanding of the 
relationships among different components \cite{girvan2002,ravasz2002}. 

A complex network could be represented by a graph. Each component of the 
network is mapped to a vertex (node), and the interaction between two
components is signified by an edge between the two corresponding nodes,
whose weight is related to the interaction strength. The challenge is to
dissect this graph based on its connection pattern. We know that to
partition a graph into two equally sized subgroups such that the number of
edges in between reaches the absolute minimum is already a NP-complete
problem, a solution is not guaranteed to be found easily; however it
is still a well-defined question. On the other hand, the question $``$How
many subgroups should a graph be divided into and how?'' is ill-posed, as we
do not have an objective function to optimize; and we have to rely on 
heuristic reasoning to proceed.

If we are interested in identifying just one community that is associated
with a specified node, the maximum flow method \cite{ford1962} turns out to
be efficient. Recently, it is applied to identifying communities of 
Internet webpages \cite{flake2000}. An community thus uncovered is
usually very small; and for this method to work well one needs {\it a priori}
knowledge of the network to select the source and sink nodes properly.
Another elegant method is based on the concept of edge betweenness
\cite{freeman1977}. The degree of betweenness of a edge is defined as
the total number of shortest paths between pair of nodes which pass through
it. By removing recursively the current edge with the highest degree
of betweenness, one expects the connectivity of the network to decrease the
most efficiently and minimal cutting operations is needed to separate the
network into subgroups \cite{girvan2002}. This idea of Girvan and Newman
\cite{girvan2002} could be readily extended to weighted graphs by assigning
each edge a length equalling its reciprocal weight. Furthermore, in the
sociology literature, there is a relatively long tradition in identifying
communities based on the criteria of reachability and shortest distance
(see, e.g., \cite{wasserman1994}).

In this paper, a new method of network community identification is
described. It is based on the concept of network Brownian motion: If an
intelligent Brownian particle lives in a given network for a long time,what
might be its perspective of the network's landscape? We suggest that,without
the need to removing edges from the network, the node-node distances
$``$measured'' by this Brownian particle can be used to construct the
community structure and to identify the central node of each community.
This idea is tested on several social and biological networks and satisfiable
results are obtained. Several ways are discussed to extend and improve 
our method.

Consider a connected network of $N$ nodes and $M$ edges. Its node set is
denoted by $V=\{1,\cdots,N\}$ and its connection pattern is specified by the
generalized adjacency matrix $A$. If there is no edge between node $i$ and
node $j$, $A_{ij}=0$; if there is an edge in between, 
$A_{ij}\equiv A_{ji}>0$ and its value signifies the interaction strength
(self-connection is allowed).
The set of nearest-neighbors of node $i$ is denoted by $E_i$. 
A Brownian particle keeps moving on the network, and at each
time step it jumps from its present position (say $i$) to a 
nearest-neighboring position ($j$). When no additional knowledge about the
network is known, it is natural to assume the following jumping probability
$P_{ij}=A_{ij}/\sum_{l=1}^N A_{il}$ (the corresponding  matrix $P$ is called
the transfer matrix). One verifies that at time $t\gg M$ the probability
$\rho(k)$ for the Brownian particle to be at any node $k$ is 
{\em nonvanishing} and equals to $\sum_{l} A_{kl}/\sum_{m,n} A_{mn}$,
proportional to the total interaction capacity $\sum_{l} A_{kl}$ of node $k$.

Define the node-node {\em distance} $d_{i,j}$ from $i$ to $j$ as the average
number of steps needed for the Brownian particle to move from $i$ through the
the network to $j$.  From some simple linear-algebra calculation 
\cite{kolman1986} it is easy to see that
\begin{equation}
d_{i,j}=\sum_{l=1}^N \left({1 \over I-B(j)}\right)_{il},
\label{eq:d_ij}
\end{equation}
where $I$ is the $N\times N$ identity matrix, and matrix $B(j)$ equals to
the transfer matrix $P$ except that $B_{lj}(j)\equiv 0$ for any $l\in V$.
The distances from all the nodes in $V$ to node $j$ can thus be obtained by
solving the linear algebraic equation 
$[I - B(j) ] \{d_{1,j},\cdots,d_{N,j}\}^T=\{1,\cdots,1\}^T$.
We are mainly interested in sparse networks with $M=O(N)$; for such networks
there exist very efficient algorithms \cite{tewarson1973,davis1999} to
calculate the root of this equation. If node $j$ has the property that
$d_{i,j}\leq d_{i,k}$ for any $k\in V$, then $j$ is tagged as a
{\em global attractor} of node $i$ ($i$ is closest to $j$ in the sense of
average distance). Similarly, if $j\in E_i$ and $d_{i,j}\leq d_{i,l}$ for any
$l\in E_i$, then $j$ is an {\em local attractor} of $i$ ($i$ is closest to
$j$ among all its nearest-neighbors). We notice that, in general the distance
from $i$ to $j$ ($d_{i,j}$) differs from that from $j$ to $i$ ($d_{j,i}$).
Consequently, if $j$ is an attractor of $i$, node $i$ is not
necessarily also an attractor of $j$.

If a graph is divided into different subgroups, on the local scale we
intuitively expect that each node $i$ will have a high probability to be in
the same subgroup as its local attractor $j$, since among all the
nearest-neighboring nodes in $E_i$, node $j$ has the shortest $``$distance''
from node $i$. For simplicity let us just {\em assume} this probability to be
unity (a possible improvement is discussed later).
Thus, we can define a  
{\em local-attractor-based community} (or simply a $``$L-community'')
as a set of nodes $L=\{i_1,\cdots,i_m\}$ such that (1) if node $i\in L$ and
node $j$ is an local attractor of $i$, then $j\in L$, (2) if $i \in L$ and 
node
$k$ has $i$ as its local attractor,then $k\in L$, and (3) any subset of $L$ is
not a L-community. Clearly, two L-communities $L_a$ and $L_b$ are either
identical ($L_a \equiv L_b$) or disjoint ($L_a \cap L_b=\emptyset$).
Based on each node's local attractor the graph could be decomposed into
a set of L-communities. 

According to  the same intuitive argument, 
on the global scale we expect that each
node will have a high probability to be in the same community as its global
attractor, and if assume this probability to be unity we can similarly
construct the {\em global-attractor-based communities} ($``$G-communities'') 
based on the global-attractor of each node. For small networks, we expect
the L- and G-community structures to be identical; while
for large networks, each G-community may contain several L-communities as its
subgroups. A community could be characterized by its size $N_c$ and
an instability index $I_c$. A node $i$ in community $C$ is referred to as
unstable if its total direct interaction with nodes in any another community
$C^\prime$, $\sum_{k\in C^\prime} A_{i k}$, is stronger than its total
direct interaction with other nodes in its own community,
$\sum_{k\in C\backslash i} A_{i k}$. 
$I_c$ is the total number of such nodes in each
community. We can also identify the {\em center} of a community (if it exists)
as the node that is the global attractor of itself.

\begin{figure*}
\begin{minipage}{.5\linewidth}
\vspace*{2.0cm}
\hspace*{-1.5cm}\psfig{file=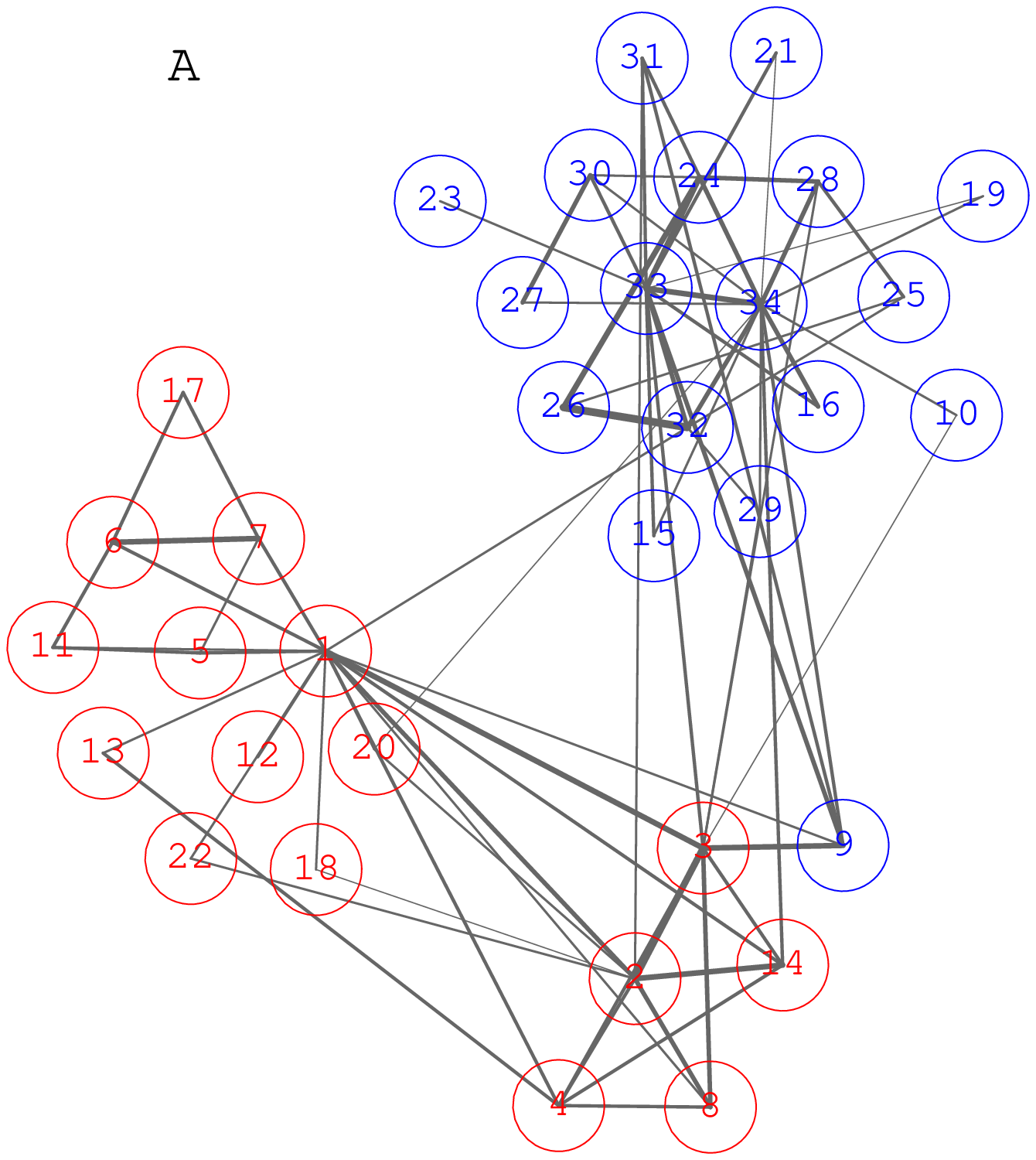,height=5.0cm}
\end{minipage}
\begin{minipage}{.5\linewidth}
\vspace*{2.0cm}
\psfig{file=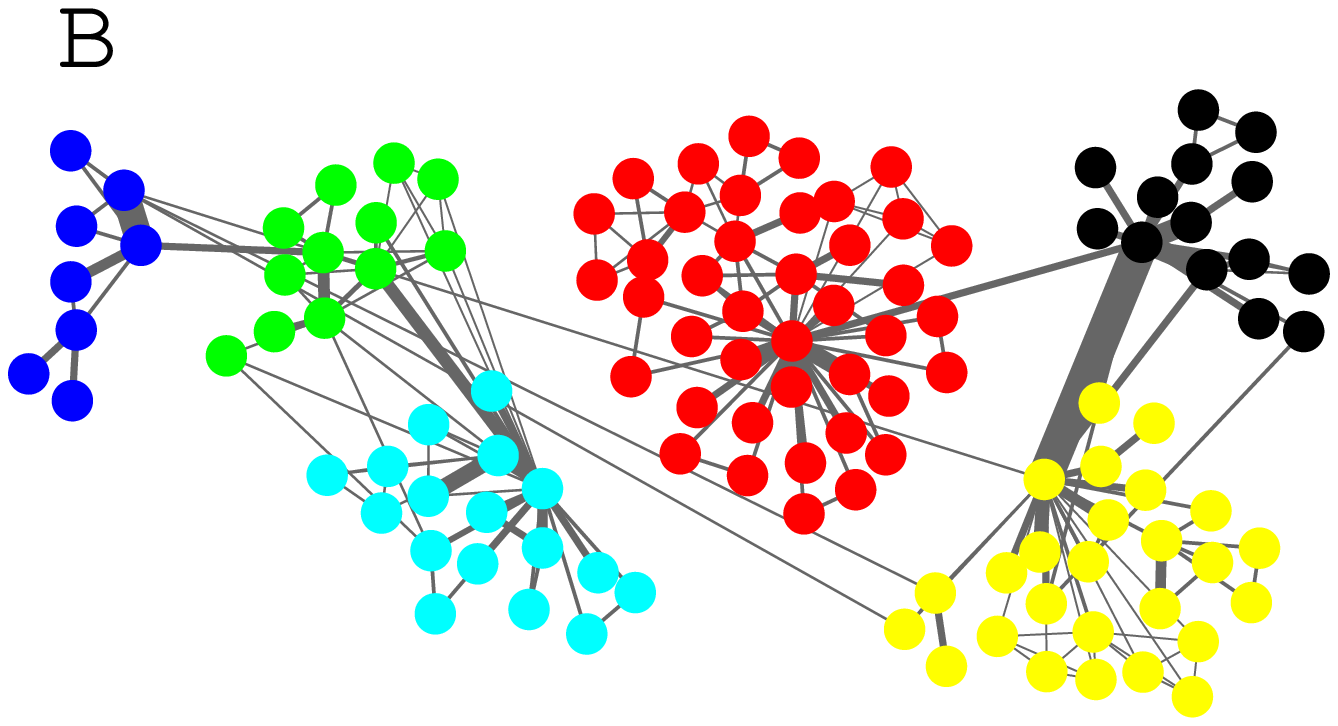,height=3.0cm}
\end{minipage}

\vspace*{1.0cm}
\hspace*{1.0cm}\psfig{file=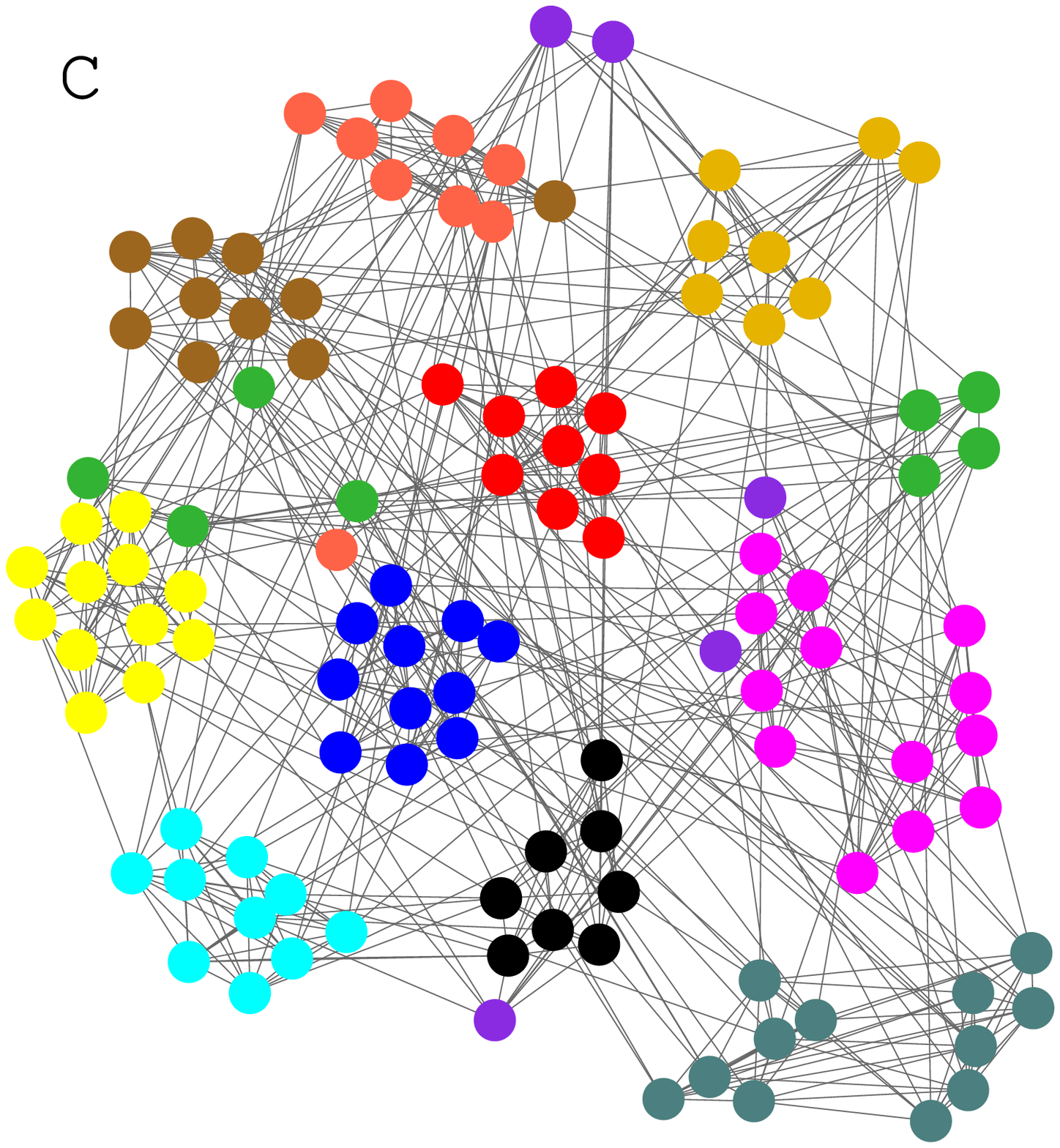,height=6.0cm}

\vspace*{1.0cm}
\hspace*{1.0cm}\psfig{file=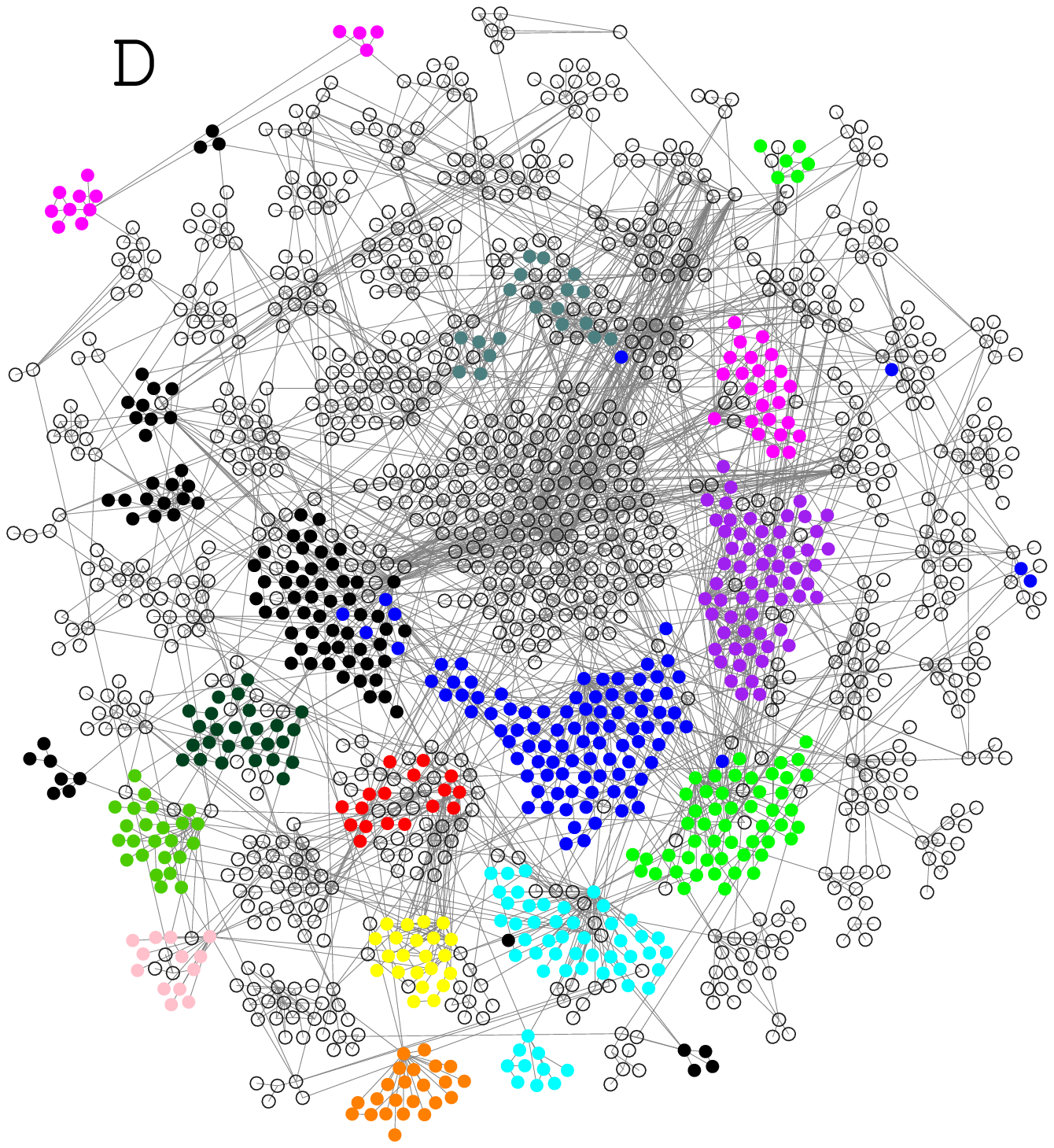,height=7.0cm}

\vskip -1.0cm
\caption{\label{fig:fig01}
(Color) Community structure of some model networks (the nodes of
the same L-community are spatially grouped together).
(a) The karate club network compiled by Zachary \cite{zachary1977} (here
nodes are colored according to their actual groupings);
(b) the scientific collaboration network compiled by Girvan and
Newman \cite{girvan2002};
(c) the foot-ball match network compiled by
Girvan and Newman \cite{girvan2002} (nodes are colored according
to their actual groupings);
and (d) the yeast protein interaction network \cite{xenarios2000,deane2002},
here nodes of the same G-community are encoded with the same color 
(open circles denote nodes in $G_1$).
}
\end{figure*}

Now we test the above-mentioned simple method on some well-documented networks
whose community structures are known. The first example is the social network
recorded by Zachary \cite{zachary1977}. 
This network contains $34$ nodes and $77$ weighted edges, and it 
was observed to spontaneously fission into two groups of size $16$ and
$18$, respectively 
\cite{zachary1977} (these two groups are marked by two colors in 
Fig.~\ref{fig:fig01}A). 
The results of our method is shown in Fig.~\ref{fig:fig01}A.
Community $L_1$ contains
$11$ elements (node $13$ is unstable and has stronger direct interaction
with $L_2$), $L_2$ has $6$ elements (node $9$ has stronger direct interaction
with $L_3$), and $L_3$ has $17$ elements.
Nodes $1$ (the manager), $3$, and $34$ (the officer) are the corresponding
centers. We find that for this network the
G-communities coincide with the L-communities.

As another example, the scientific collaboration network of 
Santa Fe Institute \cite{girvan2002} is considered. The giant connected
component contains $118$ nodes and $200$ weighted edges,
the weights are assigned according to the measure in \cite{newman2001b}. 
The present method divides the network into six L-communities, see
Fig.~\ref{fig:fig01}B. All the nodes in
 community $L_1$ (size $14$), $L_2$ ($41$), $L_4$ ($8$), $L_5$ 
($26$), and $L_6$ ($17$) are locally stable, and one node in
$L_3$ has stronger direct interaction with community $L_6$.
Same as the above example, the G-community structure is also identical
to the L-community structure.
Girvan and Newman divided this network 
into four major groups by recursively removing edges of highest degree of 
betweenness \cite{girvan2002}: the largest of which was further divided into
three subgroups and the second largest was divided into two subgroups.
There are still some minor differences between 
the six subgroups obtained by the present method and those obtained in
\cite{girvan2002}, which may be attributed 
to the fact that, in the treatment of \cite{girvan2002} the network
was regarded as unweighted.

The method is further tested on a relatively more complicated case,
the foot-ball match network compiled by Girvan and Newman \cite{girvan2002}. 
It contains $115$ nodes and $613$ unweighted edges. These $115$ teams were 
distributed into $12$ conferences by the game organizers. Based on the
connection pattern, the present method divides them
into $15$ L-communities, of which $11$ are locally stable:
$L_2$ (size $9$), $L_3$ ($13$), $L_4$ ($14$), $L_5$ ($10$), 
$L_6$ ($8$), $L_7$ ($6$), $L_8$ ($7$), $L_9$ ($6$), 
$L_{10}$ ($4$), $L_{11}$ ($6$),
and $L_{13}$ (size $9$). 
One element of $L_1$ (size $9$) has stronger interaction
with $L_{10}$, and one element of $L_{12}$ (size $10$) has stronger 
interaction
with $L_3$, and all 
the elements of $L_{14}$ (size $2$) and $L_{15}$ (size $2$)
are locally unstable. The G-communities of this network are also identical
to the L-communities. In Fig.~\ref{fig:fig01}C the community structure of
this network is shown, where nodes belonging to each identified community
are located together, and the different colors encode the actual
$12$ conferences \cite{girvan2002}. 
Figure ~\ref{fig:fig01}C indicates that the predicted communities coincide
very well with the actual communities. The community structure  obtained
by the present method is also in very good correspondence with that
obtained by Girvan and Newman \cite{girvan2002} based on edge betweenness. 

The above-studied networks all have relatively small network sizes and
the identified G-communities coincide with the L-communities.  Now we
 apply our method to the protein interaction network (yeast core 
\cite{xenarios2000,deane2002}) of baker's yeast. The giant connected 
component of this network contains $1471$ proteins and $2770$ edges 
(assumed to be unweighted, since the interaction strengths between the 
proteins are generally undetermined). 
The present method dissect this giant component into $14$ G-communities
(Table.~\ref{tab:tab01}) and into $69$ L-communities
($11$ of them contain one locally unstable node, $15$ of them
have $2$-$7$ locally unstable nodes, all the others are stable). 
The relationship
between the G- and L-communities is demonstrated in Fig~\ref{fig:fig01}D,
where proteins are grouped into L-communities and those of the same
G-community have the same color.
We see from Fig.~\ref{fig:fig01}D that if two nodes are in the same
L-community, they are very probable to be in the same G-community.
The largest G-community ($G_1$) contains more than half of the proteins
and is centered around nucleoporin {\tt YMR047C}, which, 
according to SWISS-PROT description \cite{bairoch2000}, is
$``$an essential component of  nuclear pore complex'' and $``$may be 
involved in both binding and translocation
of the proteins during nucleocytoplasmic transport''. {\tt YMR047C} 
interact directly only with $39$ other proteins (it is even not
the most connected node in the system), but associated with it is a 
group of $935$ proteins as suggested by the present method.
The protein interaction network may be evolved to facilitate efficient 
protein transportation by protein-mediated indirect interactions. 

What will happen if the protein {\tt YMR047C} is removed from the network?
The resulting perturbed system has
$1463$ nodes and $2729$ edges, and we find that its L-community structure
does not change much. 
Altogether $72$ L-communities are identified, and most of them contain more
or less the same set of elements as in the unperturbed network. However,
there is a dramatic change in the G-community structure. There are now
$21$ G-communities (the largest of which has $574$ proteins), while $G_1$
of the original system breaks up into eight smaller G-communities.
It was revealed that the most highly connected proteins in the cell are
the most important for its survival, and mutations in these proteins are 
usually lethal \cite{jeong2001}. Our work suggests that, these highly
connected proteins are especially important because they help integrating
many small functional modules (L-communities) into a larger 
unit (G-community), enabling the cell to perform concerted reactions in
response to environment stimuli.

In the above examples, the network studied are all from real-world. We 
have also tested the performance of our method to some artificial
networks generated by computer. To compare with the result of Ref.~\cite{girvan2002},
we generated an ensemble of random graphs with $128$ vertices. These
vertices are divided into four groups of $32$ vertices each. Each vertex
has on average $16$ edges, $z_{\rm out}$ of which are to vertices of
other groups, and the remaining are to vertices within its group; all
these edges are drawn randomly and independently in all the other
means. Using the method of Girvan and Newman, it was reported \cite{girvan2002}
that when $z_{\rm out}< 6$ all the vertices could be classified
with high probability. Our present method in its simplest form
could work perfectly only when $z_{\rm out}< 2.5$. In the artificial
network, the vertices are identical with each other in the statistical
sense and there is no correlation between the degrees of two neighboring
edges. Our method seems not to be the best for such kind of random
networks.

In summary, we have suggested a simple way of grouping a graph of 
nodes and edges into	
different subgraphs based on the node-node distance measured by a Brownian
particle. The basic idea was applied to several real networked systems
and very encouraging results were obtained. The concept of random walking 
was also used in some recent efforts to facilitate searching on networks
(see, e.g., \cite{tadic2002,guimera2002}), the present 
work may be the first attempt in applying it on identifying	
network community structure. Some possible extensions of our method are
immediately conceivable: First, in the present work we have assumed that
a node will be in the same community as its attractor with probability $1$.
Naturally, we can introduce a $``$inverse temperature'' $\beta$ and suppose
that node $i$ be in the same community as node $j$ with probability 
proportional to $\exp(-\beta d_{i,j})$. The present work discusses
just the zero 
temperature limit. We believe that the communities identified at zero
temperature will persist until the temperature is high enough.
Second, we can construct a gross-grained network by regarding 
each L-community as a single node, and defining the distance from
one L-community to another as the average node-node distance
between nodes in these two communities. The present method can then
be applied, and the relationship between different L-communities can
be better understood. Third, for very large networks, it is impractical
to consider the whole network when calculating node-node distance. Actually
this is not necessary, since the length of the  shortest path
between a given node and its attractor should be small. We can therefore
focus on a localized region of the network to identify 
the attractor of a given node. 

Furthermore, based on the distance measure of the present paper,
we can define a quantity called the {\em dissimilarity index} for any
two nearest-neighboring nodes. Nearest-neighboring vertices of the same
community tend to have small dissimilarity index, while those
belonging to different communities  tend to have high
dissimilarity index.  Extensions of the present work 
will be reported in a forthcoming paper
\cite{zhou2003}.

An interesting task is to use 
extended versions of the present method to explore the landscape of the 
Internet's autonomous system \cite{faloutsos1999} and that of the
metabolic network of {\it E. coli} \cite{selkov1998,ravasz2002}.

I am grateful to  M. Girvan and M. E. J. Newman for sharing data and to 
Professor R. Lipowsky for support.

\newpage

\begin{thebibliography}{23}
\expandafter\ifx\csname natexlab\endcsname\relax\def\natexlab#1{#1}\fi
\expandafter\ifx\csname bibnamefont\endcsname\relax
  \def\bibnamefont#1{#1}\fi
\expandafter\ifx\csname bibfnamefont\endcsname\relax
  \def\bibfnamefont#1{#1}\fi
\expandafter\ifx\csname citenamefont\endcsname\relax
  \def\citenamefont#1{#1}\fi
\expandafter\ifx\csname url\endcsname\relax
  \def\url#1{\texttt{#1}}\fi
\expandafter\ifx\csname urlprefix\endcsname\relax\def\urlprefix{URL }\fi
\providecommand{\bibinfo}[2]{#2}
\providecommand{\eprint}[2][]{\url{#2}}

\bibitem[{\citenamefont{Watts and Strogatz}(1998)}]{watts1998}
\bibinfo{author}{\bibfnamefont{D.~J.} \bibnamefont{Watts}} \bibnamefont{and}
  \bibinfo{author}{\bibfnamefont{S.~H.} \bibnamefont{Strogatz}},
  \bibinfo{journal}{Nature (London)} \textbf{\bibinfo{volume}{393}},
  \bibinfo{pages}{440} (\bibinfo{year}{1998}).

\bibitem[{\citenamefont{Barab\'{a}si and Albert}(1999)}]{barabasi1999}
\bibinfo{author}{\bibfnamefont{A.-L.} \bibnamefont{Barab\'{a}si}}
  \bibnamefont{and} \bibinfo{author}{\bibfnamefont{R.}~\bibnamefont{Albert}},
  \bibinfo{journal}{Science} \textbf{\bibinfo{volume}{286}},
  \bibinfo{pages}{509} (\bibinfo{year}{1999}).

\bibitem[{\citenamefont{Faloutsos et~al.}(1999)\citenamefont{Faloutsos,
  Faloutsos, and Faloutsos}}]{faloutsos1999}
\bibinfo{author}{\bibfnamefont{M.}~\bibnamefont{Faloutsos}},
  \bibinfo{author}{\bibfnamefont{P.}~\bibnamefont{Faloutsos}},
  \bibnamefont{and}
  \bibinfo{author}{\bibfnamefont{C.}~\bibnamefont{Faloutsos}},
  \bibinfo{journal}{Comput.\ Commun.\ Res.} \textbf{\bibinfo{volume}{29}},
  \bibinfo{pages}{251} (\bibinfo{year}{1999}).

\bibitem[{\citenamefont{Strogatz}(2001)}]{strogatz2001}
\bibinfo{author}{\bibfnamefont{S.~H.} \bibnamefont{Strogatz}},
  \bibinfo{journal}{Nature (London)} \textbf{\bibinfo{volume}{410}},
  \bibinfo{pages}{268} (\bibinfo{year}{2001}).

\bibitem[{\citenamefont{Albert and Barab\'{a}si}(2002)}]{albert2002}
\bibinfo{author}{\bibfnamefont{R.}~\bibnamefont{Albert}} \bibnamefont{and}
  \bibinfo{author}{\bibfnamefont{A.-L.} \bibnamefont{Barab\'{a}si}},
  \bibinfo{journal}{Rev.\ Mod.\ Phys.} \textbf{\bibinfo{volume}{74}},
  \bibinfo{pages}{47} (\bibinfo{year}{2002}).

\bibitem[{\citenamefont{Dorogovtsev and Mendes}(2002)}]{dorogovtsev2002}
\bibinfo{author}{\bibfnamefont{S.~N.} \bibnamefont{Dorogovtsev}}
  \bibnamefont{and} \bibinfo{author}{\bibfnamefont{J.~F.~F.}
  \bibnamefont{Mendes}}, \bibinfo{journal}{Adv.\ Phys.}
  \textbf{\bibinfo{volume}{51}}, \bibinfo{pages}{1079} (\bibinfo{year}{2002}).

\bibitem[{\citenamefont{Girvan and Newman}(2002)}]{girvan2002}
\bibinfo{author}{\bibfnamefont{M.}~\bibnamefont{Girvan}} \bibnamefont{and}
  \bibinfo{author}{\bibfnamefont{M.~E.~J.} \bibnamefont{Newman}},
  \bibinfo{journal}{Proc.\ Natl.\ Acad.\ Sci.\ U.S.A.}
  \textbf{\bibinfo{volume}{99}}, \bibinfo{pages}{7821} (\bibinfo{year}{2002}).

\bibitem[{\citenamefont{Ravasz et~al.}(2002)\citenamefont{Ravasz, Somera,
  Mongru, Oltvai, and Barab\'{a}basi}}]{ravasz2002}
\bibinfo{author}{\bibfnamefont{E.}~\bibnamefont{Ravasz}},
  \bibinfo{author}{\bibfnamefont{A.~L.} \bibnamefont{Somera}},
  \bibinfo{author}{\bibfnamefont{D.~A.} \bibnamefont{Mongru}},
  \bibinfo{author}{\bibfnamefont{Z.~N.} \bibnamefont{Oltvai}},
  \bibnamefont{and} \bibinfo{author}{\bibfnamefont{A.-L.}
  \bibnamefont{Barab\'{a}basi}}, \bibinfo{journal}{Science}
  \textbf{\bibinfo{volume}{297}}, \bibinfo{pages}{1551} (\bibinfo{year}{2002}).

\bibitem[{\citenamefont{Ford and Fulkerson}(1979)}]{ford1962}
\bibinfo{author}{\bibfnamefont{L.}~\bibnamefont{Ford}} \bibnamefont{and}
  \bibinfo{author}{\bibfnamefont{D.}~\bibnamefont{Fulkerson}},
  \emph{\bibinfo{title}{Flows in networks}} (\bibinfo{publisher}{Princeton
  University Press}, \bibinfo{address}{Princeton, New Jesey},
  \bibinfo{year}{1979}).

\bibitem[{\citenamefont{Flake et~al.}(2000)\citenamefont{Flake, Lawrence, and
  Giles}}]{flake2000}
\bibinfo{author}{\bibfnamefont{G.~W.} \bibnamefont{Flake}},
  \bibinfo{author}{\bibfnamefont{S.}~\bibnamefont{Lawrence}}, \bibnamefont{and}
  \bibinfo{author}{\bibfnamefont{C.~L.} \bibnamefont{Giles}}, in
  \emph{\bibinfo{booktitle}{Proceedings of the Sixth International Conference
  on Knowledge Discovery and Data Mining (ACM SIGKDD-2000)}}
  (\bibinfo{year}{2000}), pp. \bibinfo{pages}{150--160}.

\bibitem[{\citenamefont{Freeman}(1977)}]{freeman1977}
\bibinfo{author}{\bibfnamefont{L.~C.} \bibnamefont{Freeman}},
  \bibinfo{journal}{Sociometry} \textbf{\bibinfo{volume}{40}},
  \bibinfo{pages}{35} (\bibinfo{year}{1977}).

\bibitem[{\citenamefont{Wasserman and Faust}(1994)}]{wasserman1994}
\bibinfo{author}{\bibfnamefont{S.}~\bibnamefont{Wasserman}} \bibnamefont{and}
  \bibinfo{author}{\bibfnamefont{K.}~\bibnamefont{Faust}},
  \emph{\bibinfo{title}{Social Network Analysis: Methods and Applications}}
  (\bibinfo{publisher}{Cambridge University Press}, \bibinfo{address}{UK},
  \bibinfo{year}{1994}).

\bibitem[{\citenamefont{Kolman}(1973)}]{kolman1986}
\bibinfo{author}{\bibfnamefont{B.}~\bibnamefont{Kolman}},
  \emph{\bibinfo{title}{Elementary Linear
Algebra (4th Edition)}} (\bibinfo{publisher}{MacMillan Publisher},
  \bibinfo{address}{HB}, \bibinfo{year}{1986}).

\bibitem[{\citenamefont{Tewarson}(1973)}]{tewarson1973}
\bibinfo{author}{\bibfnamefont{R.~P.} \bibnamefont{Tewarson}},
  \emph{\bibinfo{title}{Sparse Matrices}} (\bibinfo{publisher}{Academic Press},
  \bibinfo{address}{New York}, \bibinfo{year}{1973}).

\bibitem[{\citenamefont{Davis and Duff}(1999)}]{davis1999}
\bibinfo{author}{\bibfnamefont{T.~A.} \bibnamefont{Davis}} \bibnamefont{and}
  \bibinfo{author}{\bibfnamefont{I.~S.} \bibnamefont{Duff}},
  \bibinfo{journal}{ACM Trans.\ Math.\ Software} \textbf{\bibinfo{volume}{25}},
  \bibinfo{pages}{1} (\bibinfo{year}{1999}).

\bibitem[{\citenamefont{Zachary}(1977)}]{zachary1977}
\bibinfo{author}{\bibfnamefont{W.~W.} \bibnamefont{Zachary}},
  \bibinfo{journal}{J.\ Anthropol.\ Res.} \textbf{\bibinfo{volume}{33}},
  \bibinfo{pages}{452} (\bibinfo{year}{1977}).

\bibitem[{\citenamefont{Newman}(2001)}]{newman2001b}
\bibinfo{author}{\bibfnamefont{M.~E.~J.} \bibnamefont{Newman}},
  \bibinfo{journal}{Phys.\ Rev.\ E} \textbf{\bibinfo{volume}{64}},
  \bibinfo{pages}{016132} (\bibinfo{year}{2001}).

\bibitem[{\citenamefont{Xenarios et~al.}(2000)\citenamefont{Xenarios, Rice,
  Salwinski, Baron, Marcotte, and Eisenberg}}]{xenarios2000}
\bibinfo{author}{\bibfnamefont{I.}~\bibnamefont{Xenarios}},
  \bibinfo{author}{\bibfnamefont{D.~W.} \bibnamefont{Rice}},
  \bibinfo{author}{\bibfnamefont{L.}~\bibnamefont{Salwinski}},
  \bibinfo{author}{\bibfnamefont{M.~K.} \bibnamefont{Baron}},
  \bibinfo{author}{\bibfnamefont{E.~M.} \bibnamefont{Marcotte}},
  \bibnamefont{and}
  \bibinfo{author}{\bibfnamefont{D.}~\bibnamefont{Eisenberg}},
  \bibinfo{journal}{Nucleic Acids Res.} \textbf{\bibinfo{volume}{28}},
  \bibinfo{pages}{289} (\bibinfo{year}{2000}).

\bibitem[{\citenamefont{Deane et~al.}(2002)\citenamefont{Deane, Salwinski,
  Xenarios, and Eisenberg}}]{deane2002}
\bibinfo{author}{\bibfnamefont{C.~M.} \bibnamefont{Deane}},
  \bibinfo{author}{\bibfnamefont{L.}~\bibnamefont{Salwinski}},
  \bibinfo{author}{\bibfnamefont{I.}~\bibnamefont{Xenarios}}, \bibnamefont{and}
  \bibinfo{author}{\bibfnamefont{D.}~\bibnamefont{Eisenberg}},
  \bibinfo{journal}{Mol.\ Cell.\ Proteomics} \textbf{\bibinfo{volume}{1}},
  \bibinfo{pages}{349} (\bibinfo{year}{2002}).

\bibitem[{\citenamefont{Bairoch and Apweiler}(2000)}]{bairoch2000}
\bibinfo{author}{\bibfnamefont{A.}~\bibnamefont{Bairoch}} \bibnamefont{and}
  \bibinfo{author}{\bibfnamefont{R.}~\bibnamefont{Apweiler}},
  \bibinfo{journal}{Necleic Acids Res.} \textbf{\bibinfo{volume}{28}},
  \bibinfo{pages}{45} (\bibinfo{year}{2000}).

\bibitem[{\citenamefont{Jeong et~al.}(2001)\citenamefont{Jeong, Mason,
  Barab\'{a}si, and Oltvai}}]{jeong2001}
\bibinfo{author}{\bibfnamefont{H.}~\bibnamefont{Jeong}},
  \bibinfo{author}{\bibfnamefont{S.~P.} \bibnamefont{Mason}},
  \bibinfo{author}{\bibfnamefont{A.-L.} \bibnamefont{Barab\'{a}si}},
  \bibnamefont{and} \bibinfo{author}{\bibfnamefont{Z.~N.}
  \bibnamefont{Oltvai}}, \bibinfo{journal}{Nature (London)}
  \textbf{\bibinfo{volume}{411}}, \bibinfo{pages}{41} (\bibinfo{year}{2001}).

\bibitem[{\citenamefont{Tadi\'{c}}(2002)}]{tadic2002}
\bibinfo{author}{\bibfnamefont{B.}~\bibnamefont{Tadi\'{c}}},
  \bibinfo{journal}{Eur. Phys. J. B} \textbf{\bibinfo{volume}{23}},
  \bibinfo{pages}{221} (\bibinfo{year}{2002}).

\bibitem[{\citenamefont{Guimer\`{a} et~al.}(2002)\citenamefont{Guimer\`{a},
  Diaz-Guilera, Vega-Redondo, Cabrales, and Arenas}}]{guimera2002}
\bibinfo{author}{\bibfnamefont{R.}~\bibnamefont{Guimer\`{a}}},
  \bibinfo{author}{\bibfnamefont{A.}~\bibnamefont{Diaz-Guilera}},
  \bibinfo{author}{\bibfnamefont{F.}~\bibnamefont{Vega-Redondo}},
  \bibinfo{author}{\bibfnamefont{A.}~\bibnamefont{Cabrales}}, 
\bibnamefont{and}
  \bibinfo{author}{\bibfnamefont{A.}~\bibnamefont{Arenas}},
  \bibinfo{journal}{Phys.\ Rev.\ Lett.} \textbf{\bibinfo{volume}{89}},
  \bibinfo{pages}{248701} (\bibinfo{year}{2002}).

\bibitem{zhou2003}
H.~Zhou (2003), in preparation.

\bibitem[{\citenamefont{Selkov Jr et~al.}(1998)\citenamefont{Selkov Jr, 
Grechkin, Mikhailova,  and Selkov}}]{selkov1998}
\bibinfo{author}{\bibfnamefont{E.}~\bibnamefont{Selkov Jr}},
  \bibinfo{author}{\bibfnamefont{Y.}~\bibnamefont{Grechkin}},
  \bibinfo{author}{\bibfnamefont{N.}~\bibnamefont{Mikhailova}},
  \bibnamefont{and} \bibinfo{author}{\bibfnamefont{E.}~\bibnamefont{Selkov}},
  \bibinfo{journal}{Nucleic Acid Res.} \textbf{\bibinfo{volume}{26}},
  \bibinfo{pages}{43} (\bibinfo{year}{1998}).

\end{thebibliography}

\newpage

\begin{table}
\caption{\label{tab:tab01}
G-communities of yeast's protein interaction 
network \cite{xenarios2000,deane2002}. $N_c$ is the community
size, $I_c$ is the number of locally unstable nodes.}
\begin{tabular}{llll||llll}
index &$N_c$  &$I_c$  &center        &index    &$N_c$ &$I_c$ &center \\ \hline
$G_1$ &$935$  &$7$    &{\tt YMR047C} &$G_8$    &$52$  &$1$   &{\tt YBR109C} \\
$G_2$ &$90$   &$3$    &{\tt YNL189W} &$G_9$    &$37$  &$1$   &{\tt YGR218W} \\
$G_3$ &$17$   &$4$    &{\tt YER148W} &$G_{10}$ &$19$  &$2$   &{\tt YML109W} \\
$G_4$ &$57$   &$5$    &{\tt YFL039C} &$G_{11}$ &$26$  &$0$   &{\tt YDR167W} \\
$G_5$ &$97$   &$3$    &{\tt YDR388W} &$G_{12}$ &$24$  &$0$   &{\tt YDL140C} \\
$G_6$ &$59$   &$0$    &{\tt YJR022W} &$G_{13}$ &$13$  &$0$   &{\tt YOL051W} \\
$G_7$ &$22$   &$0$    &{\tt YDR448W} &$G_{14}$ &$23$  &$0$   &{\tt YJR091C}
\end{tabular}
\end{table}

\end{document}